\documentclass[12pt]{iopart}
\usepackage{iopams}
\usepackage{setstack}
\usepackage{lmodern} 
\usepackage{graphicx}
\usepackage{epigraph}
\usepackage{amssymb}
\usepackage{paralist}
\usepackage{color}
\usepackage{xfrac} 

\usepackage{amscd}
\usepackage{dsfont}
\usepackage{bm}

\def \eps{\varepsilon}

\def\K{\mathcal{K}}

\def\osc{ {\mathrm{osc}}}

\def\HI{\hat{H}_{\scriptscriptstyle\mathrm{I}}}
\def\HK{\hat{H}_{\scriptscriptstyle\mathrm{K}}}
\def\UI{U_{\scriptscriptstyle\mathrm{I}}}
\def\UK{U_{\scriptscriptstyle\mathrm{K}}}

\def\trace{\Tr}

\def\Re{ {\mathrm{Re}}}

\newif\ifdraft
\draftfalse

\DeclareGraphicsExtensions{.pdf, .png, .bmp, .jpg}

\newcommand{\iu}{\mathrm{i}}
\newcommand{\eu}{\mathrm{e}}

\newcommand{\eg}{{\em e.g.}\ }
\newcommand{\ie}{{\em i.e.}\ }
\newcommand{\cf}{{\em cf.}\ }
\newcommand{\teq}{\! = \!}

\newcommand{\kc}{{k_\text{cut}}}
\renewcommand{\Re}{\operatorname{Re}}

\newcommand{\ldmax}{{\tilde{\lambda}_{\text{max}}}}
\newcommand{\binom}[2]{{#1 \choose #2}}
\newcommand{\operatorname}[1]{\text{#1}}
\newcommand{\eqref}[1]{(\ref{#1})}

\begin{document}
\title[Particle-Time Duality II]{Particle-Time Duality in the Kicked Ising Chain II: Applications to the Spectrum}

\author{M. Akila,  D. Waltner, B. Gutkin, T. Guhr}
\address{
Faculty of Physics, University Duisburg-Essen, Lotharstr. 1, 47048 Duisburg, Germany
}
\begin{abstract}
Previously, we demonstrated that the dynamics of kicked spin chains possess a remarkable duality property. The trace of the unitary evolution operator
for $N$ spins at time $T$ is related to one of a non-unitary evolution operator for $T$ spins at time $N$. Using this duality relation we obtain the oscillating part of the density of states for a large number of spins. Furthermore, the duality relation explains the anomalous  short-time behavior of the spectral form factor previously observed  in the literature.
\end{abstract}

Pacs: 02.70.Hm, 05.45.Mt, 05.45.Pq


\section{Introduction}
The correspondence between classical and quantum mechanical systems for long times, of the order of the Heisenberg time, is well understood on the single particle level, see \cite{haake,haake2,haake3,stoeckmann}. At short times, however, spectral correlations are only given by few short system specific orbits which yield non universal results \cite{Berry}. It is an interesting question whether a limit of large particle numbers, which leads to an exponentially growing number of short orbits, might bring on its own kind of universality. To address this question we introduced the notion of a dual operator for kicked systems with nearest neighbor interaction, see \cite{Gutkin,prevPap}, which allows one to calculate traces of short time evolution operators for arbitrary particle numbers. While the dimension of evolution operators \(U_N\) grows exponentially with the particle number \(N\) the dimension of the dual operator \(\tilde{U}_T\) depends only (exponentially) on the number of time steps \(T\) the kicked system evolves. 

In \cite{prevPap} we used a Kicked Ising Chain (KIC) to study the spectrum of \(\tilde{U}_T\). As it turned out it is generically non-unitary and its structure highly depends on the properties of the considered parameter regime, ranging from integrable to chaotic dynamics.
Here, we use \(\tilde{U}_T\) to determine, for large particle numbers, the asymptotic behavior of the smooth part of the spectral density and the form factor of said KIC.
In this limit we only need to calculate the eigenvalues of the finite dimensional dual operator and not of the infinite dimensional  \(U_N\). Additional simplifications stem from non-unitarity as for large \(N\) we can restrict ourselves to the largest eigenvalues of \(\tilde{U}_T\).

The outline of the paper is as follows: At first, in section \ref{sec:model}, we give a brief overview over the model and the main results from \cite{prevPap}.
In section  \ref{section3} we use the dual operator to determine the spectral density.
The spectral form factor both for short and long times \(T\) is covered in section \ref{section4} and the conclusion is given in section \ref{section5}.

\section{Model and Duality}
\label{sec:model}

The Kicked Ising Chain model (KIC), as used in \cite{prosen2,prosen2007,prosenB3-d,prosenJt-2,prevPap}, is a one dimensional periodic chain of \(N\) spin-\(1/2\) particles. Its time evolution in term of unit time steps can be described by the Floquet operator
\begin{equation}
U_N=\UI(J) \UK(b,\varphi)\,,
\end{equation}
whose separate parts correspond to the Ising part coupling nearest neighbor spins \(\sigma_n\),
\begin{equation}
\UI(J)=\exp\left(-\iu J\sum_{n=1}^N \hat{\sigma}^z_{n}\hat{\sigma}^z_{n+1}\right)\,,
\end{equation}
and a kicking part which acts locally on every spin,
\begin{equation}
\UK(b,\varphi) = \exp\left(-\iu \sum_{n=1}^N \bm{b} \cdot \bm{\hat{\sigma}}_{n}\right)\,.
\end{equation}
Herein the magnetic field \(\bm{b}\teq b\,(\sin{\varphi},\,0,\,\cos{\varphi})\) can be restricted to the \((x,z)\) plane without loss of generality and \(\bm{\hat{\sigma}}_{i}\teq(\hat{\sigma}^x_{i}, \hat{\sigma}^y_{i}, \hat{\sigma}^z_{i})\) are the Pauli matrices for spin \(i\). Due to the periodic boundary conditions we have \(\hat{\sigma}^z_{N+1}\teq\hat{\sigma}^z_{1}\). The relevant range for the parameters \(J,\,b,\,\varphi\) is \([0,\pi/2]\). The different regimes of the KIC can be distinguished by the spectral statistics of \(U_N\). For instance \(\varphi\teq 0\) leads to a magnetic field parallel to the coupling and is thus an integrable regime. On the other hand, also an orthogonal field (\(\varphi\teq \pi/2\)) leads to integrable dynamics, while chaos can be found in between these limiting cases.

Traces of the transfer matrix,
\begin{equation}
 Z(N,T)=\trace{U_N^T} \,, \label{trace}
\end{equation}
can be understood as a partition function
\begin{equation}
\fl
 Z(N,T)=\sum_{\{\sigma_{n,t}= \pm 1\}}\exp\left(-\iu\sum_{n=1}^N\sum_{t=1}^T\left(
  J\sigma_{n,t}\sigma_{n+1,t}
 +K\sigma_{n,t}\sigma_{n,t+1}
 +h\sigma_{n,t}
 +i\eta\right)\right) \label{partfunction}
\end{equation}
of a classical two dimensional Ising model with complex parameters and periodic boundary conditions in both directions. The summation covers only classical states of either spin up \((+1)\) or down \((-1) \) on each site. The new parameters in terms of the original ones are given by
\begin{equation}
{\rm e}^{-4iK}=1-\frac{1}{{x}^2}
\,,\quad
e^{4\eta}= x^2(x^2-1)
\,,\quad
e^{-2ih} =\frac{\cos b-i\sin b \cos \varphi}{\cos b+i\sin b \cos \varphi}
\,,
\label{eq:khhparam}
\end{equation}
with \(x\teq \sin{b}\sin{\varphi}\).
In the same fashion as we expanded the quantum mechanical system onto the two dimensional model we can contract it again, but we are free to choose either particle or time direction. Contraction along the former direction leads to the dual transfer operator
\begin{equation}
 \tilde{U}_T=g^T\UI(K) \UK(\tilde{b},\tilde{\varphi})\,,\label{dualmatrix}
\end{equation}
where the dual parameters are determined  by \eqref{eq:khhparam} if one exchanges \(J\leftrightarrow K\) and \((b,\varphi)\leftrightarrow(\tilde{b},\tilde{\varphi})\), \eg
\begin{equation}\label{dualrela}
\fl
\qquad {\rm e}^{-4\iu J}=1-\frac{1}{\tilde{x}^2}, \quad \tilde{x}=\sin \tilde{b}\sin \tilde{\varphi}\quad
\text{and additionally}\quad
 g^4=\frac{x^2(x^2-1)}{\tilde{x}^2(\tilde{x}^2-1)}
\,.
\end{equation}
Up to the correction factor \(g\) this dual matrix is still of the same form as \(U_N\) in the original KIC, however the parameters of the model are complex making \(\tilde{U}_T\) generically non-unitary. Traces of both transfer matrices \(U_N^T\) and \(\tilde{U}_T^N\) describe the same classical partition and thus we find the trace duality
\begin{equation}
  \trace{{U}_N^T}=Z(N,T)= \trace{\tilde{U}_T}^N\,.\label{duality}
\end{equation}
The crucial difference between both matrices is their dimension, in the original KIC it was \(2^N\!\times2^N\) but in the dual picture the role of the particle number is taken by the time steps and the dimension is \(2^T\!\times2^T\). Especially in the case of short propagation times this allows one to consider arbitrarily large particle numbers. For further detail on the spectrum of \(\tilde{U}_T\) and the mapping mechanism described above we refer to \cite{prevPap}.

\section{Density of States}\label{section3}
As a first application we look at the density of states, first for the kicked system and later on also in the limit of continuous dynamics. In the discrete case we approximate the spectral density by only the largest dual eigenvalues and find sharp transitions between the asymptotic densities for \(N\gg 1\) when varying the system parameters, see section \ref{spDeKIC}. Subsequently we look at the limit of small parameters, section \ref{smaPar}, which already approximates the time continuous limit.
In the continuous case, section \ref{cont}, we utilize the duality relation to reproduce the Gaussian behavior of the density.

\subsection{Spectral Density for the KIC}
\label{spDeKIC}
\begin{figure}[tbp]
\centering
\includegraphics[height=0.35\textwidth]{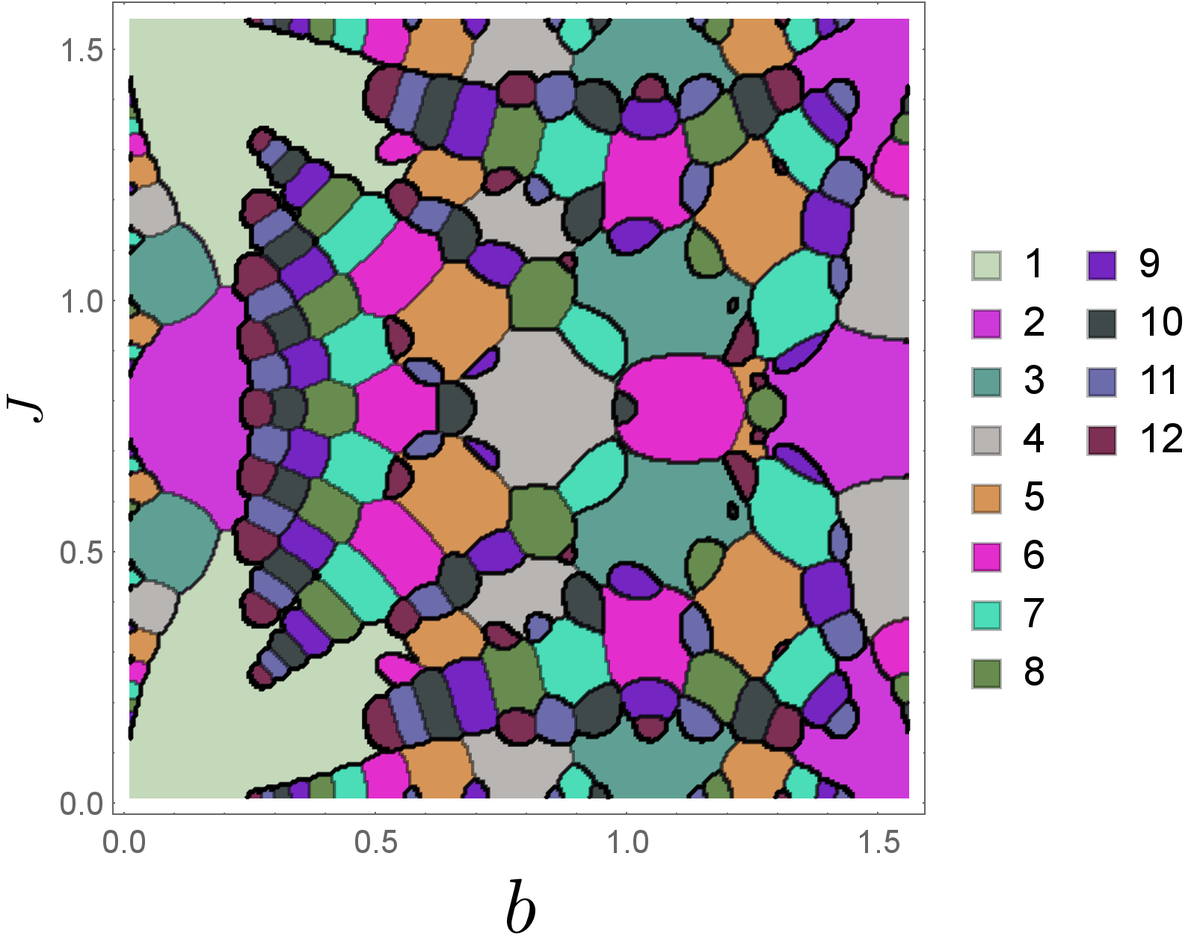}
\caption{Depicted is the time \(T_1\) for which the set of dual operators \(\tilde{U}_T\) for fixed parameters but \(1\leq T \leq 12\) contains  the eigenvalue with the largest magnitude compared to the other times.
The third system parameter \(\varphi\teq \pi/8\) is chosen such that the picture covers a large region with non-integrable system behavior.
}
\label{fig:HFMtransit1}
\end{figure}

The spectrum of the quantum evolution operator \(U_N\) comprises \(2^N\) unitary eigenvalues \(\eu^{\iu\vartheta_n}\) with quasi-energies \(\vartheta_n\).
The corresponding spectral density can be written as a Fourier series of traces of the evolution operator,
\begin{equation}
\rho{(\vartheta)}=\frac{1}{2^N}\sum_{n=1}^{2^N} \delta{(\vartheta-\vartheta_n)}
=\frac{1}{2\pi}+ \frac{1}{2^N \pi}\Re\sum_{T=1}^\infty \eu^{-\iu T\vartheta} \Tr{U_N^T}\,. \label{density_basic}
\end{equation}
We  are  interested in  the non-constant term on the right hand side of (\ref{density_basic}) representing the oscillating   part  $\rho_\osc(\vartheta)$ of the density. For large \(N\)   the traces can be approximated by powers of the eigenvalues with the largest magnitude of the dual operator,
\begin{equation}
\Tr U_N^T=\Tr \tilde{U}_T^N
\underset{N\gg 1}{\approx}
\sum_{m=1}^{m_{\text{cut}}} \tilde{\lambda}_{\text{max,}m}^N(T)\,,
\end{equation}
where   \(\tilde{\lambda}_{\text{max,}m}\)  are the eigenvalues of $\tilde{U}_T$ ordered such that they decrease in magnitude. The cut-off parameter $m_{\text{cut}}$ depends on $N$ and has to be chosen such that the error resulting from the neglected eigenvalues is small. Ordering the $|\ldmax(T)| $ for all time steps according to their absolute value,
\begin{equation}
|\ldmax(T_1)|\geq |\ldmax(T_2)|\geq |\ldmax(T_3)|\geq\dots\,,
\end{equation}
introduces a hierarchy of the corresponding points in time, $T_1,T_2 \dots $ . However, these points need not necessarily be different, for instance due to degeneracies.
Clearly, in the large \(N\) limit $\ldmax^N(T_1)$ dominates the rest of the eigenvalues and  $T_1$ determines  the asymptotic period of the $\rho_{\osc}{(\vartheta)}$  oscillations. Since  the value of  $T_1$ depends on the parameters of the KIC, the asymptotic form of $\rho_{\osc}{(\vartheta)}$ undergoes abrupt transitions  under the change of $\bm b$ and $J$. To illustrate this we color code the largest components for some parameter regime in figure \ref{fig:HFMtransit1}. Previously, similar transitions in the oscillatory behavior of $\rho{(\vartheta)}$ were observed in the  two-dimensional KIC \cite{prosen2014}.

\begin{figure}
\includegraphics[width=0.5\textwidth]{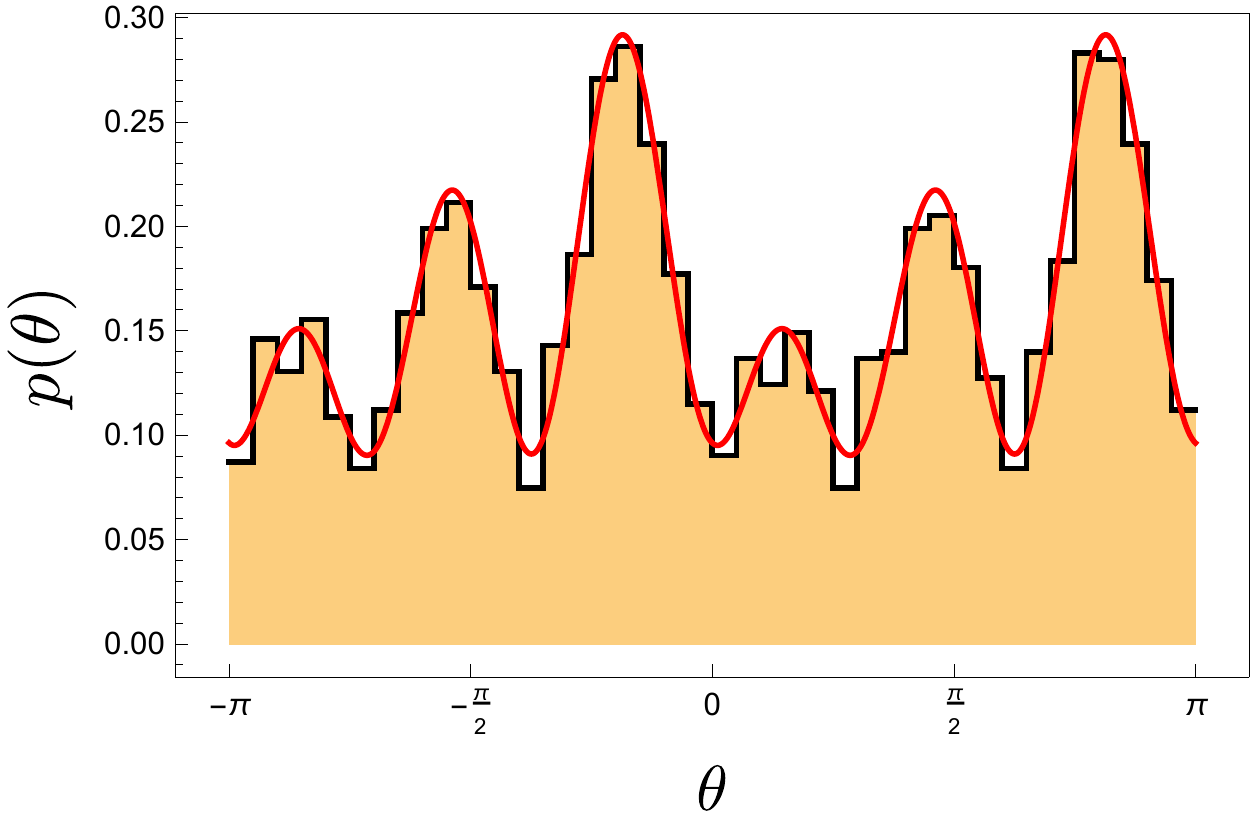}
\hfill
\includegraphics[width=0.45\textwidth]{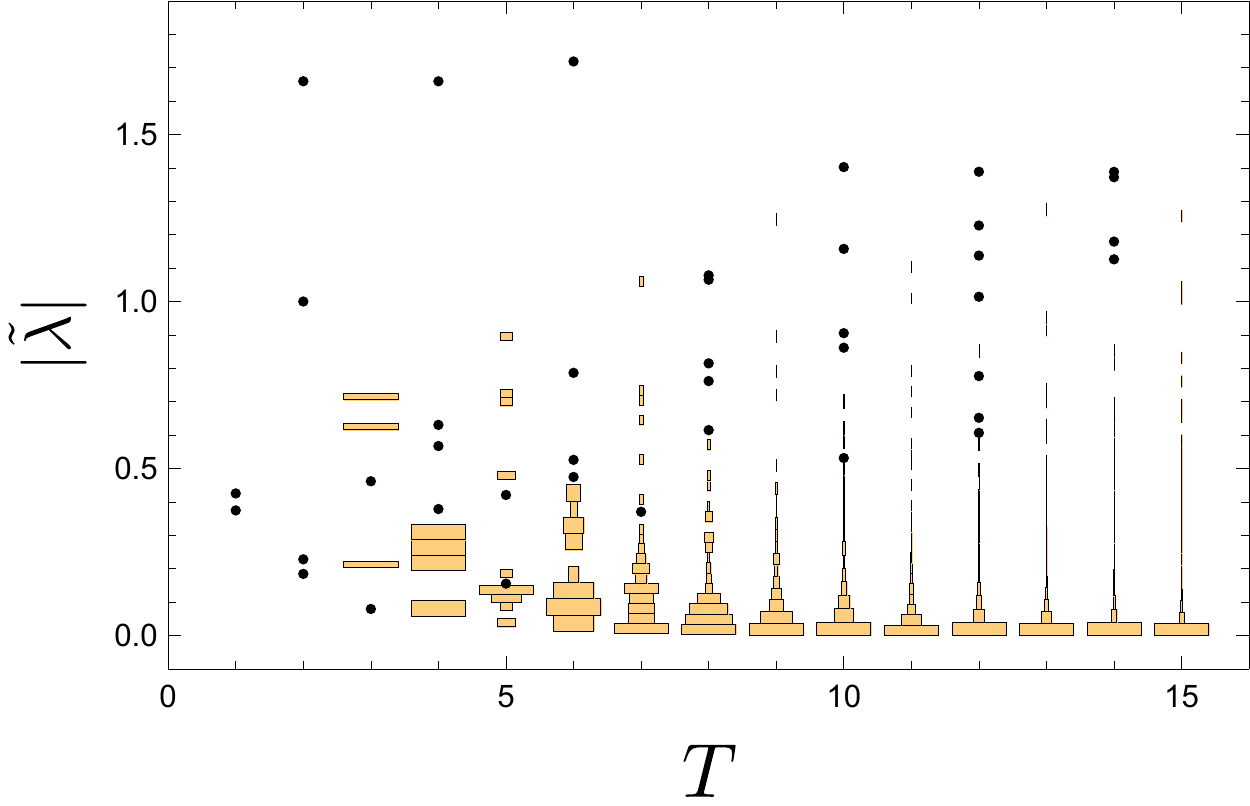}
\caption{Histogram of the spectral density (left hand side) for \(J\teq 1,\, b^x\teq 1.4,\, b^z\teq 0.4\) and \(N\teq 14\), the red line shows an approximation to the density using the 3 largest Fourier components as given by the dual spectrum. The right hand side shows histograms of the dual spectrum (absolute value) over \(T\) for the same parameter. Bar thickness corresponds to local eigenvalue density, single points to isolated eigenvalues. Clearly visible are the 3 outliers which contribute to $\rho_{\osc}(\vartheta)$.}
\label{fig:ApproximatedFourier1}
\end{figure}

\begin{figure}
\includegraphics[width=0.5\textwidth]{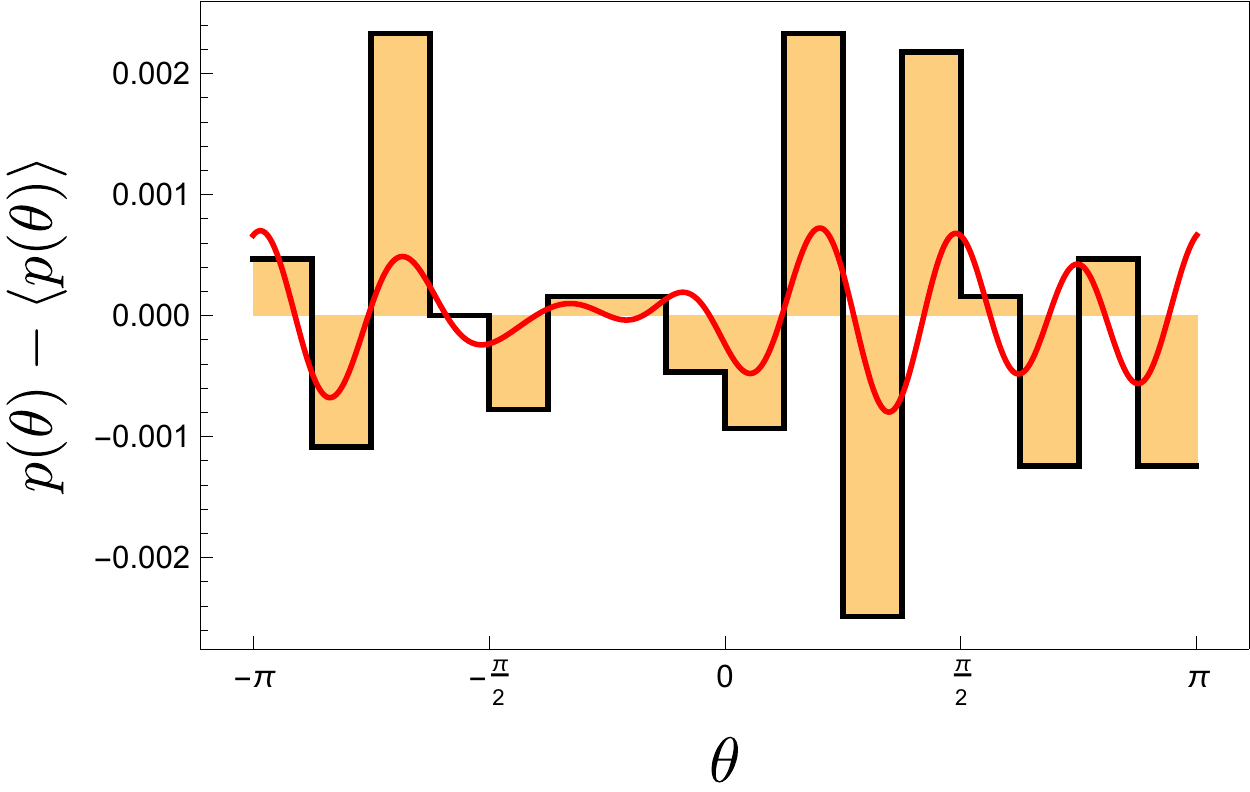}
\hfill
\includegraphics[width=0.45\textwidth]{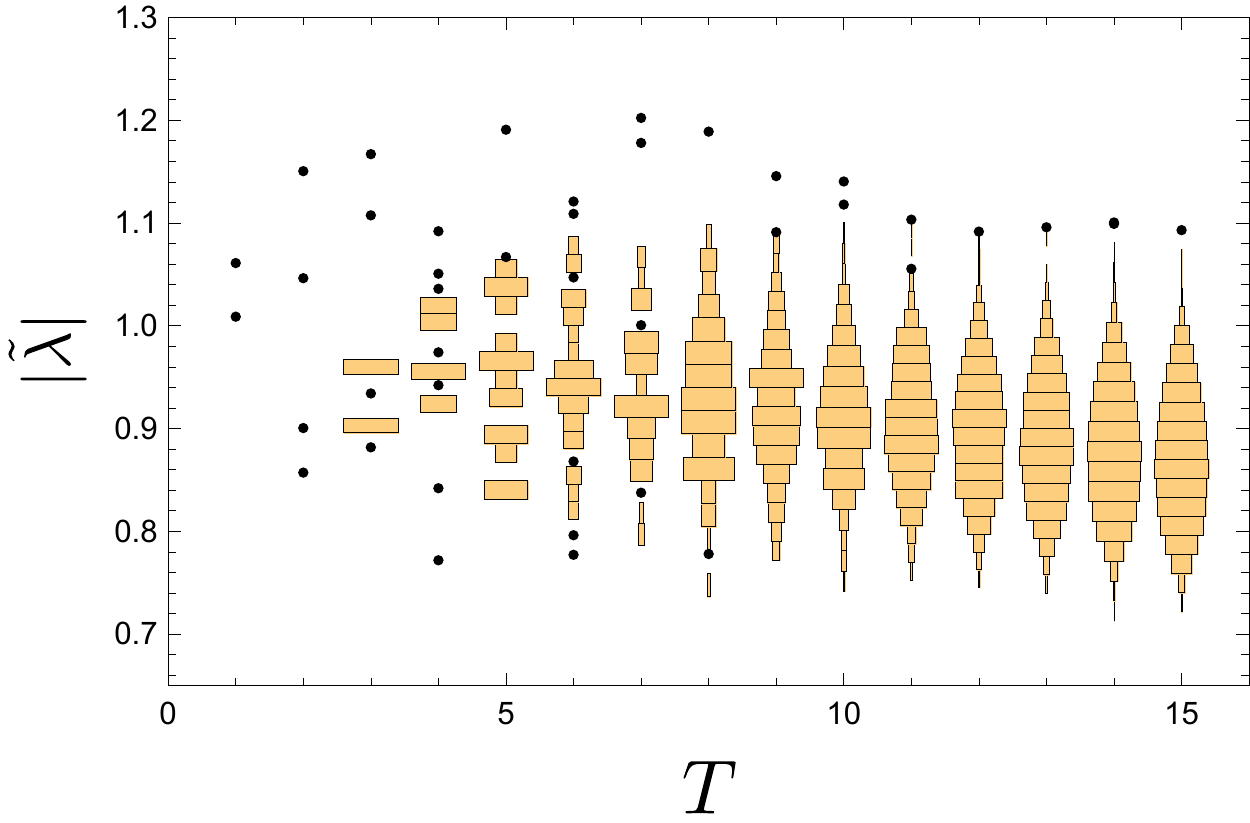}
\caption{Histogram of the spectral density (left hand side) for \(J\teq 0.7,\, b\teq 0.9 \sqrt{2},\,\varphi\teq \pi/4\) and \(N\teq 14\), the red line shows an approximation to the density using the 4 largest Fourier components as given by the dual spectrum. Due to the small size of the fluctuations only the relative change compared to the flat background is shown. The right hand side shows histograms of the dual spectrum (absolute value) in the same fashion as before in figure \ref{fig:ApproximatedFourier1}. But, in this chaotic case no clear outliers exist, making the approximation visibly less accurate.}
\label{fig:ApproximatedFourier2}
\end{figure}

For a finite $N$ 
several  values \(\ldmax(T_k)\),  $k\leq\kc$  can be used to approximate $\rho_\osc(\vartheta)$. Restricting the sum in (\ref{density_basic}) to  these few terms yields
\begin{equation}
\rho_{\osc}{(\vartheta)}\approx \frac{1}{\pi2^N} \Re \sum_{k=1}^\kc \eu^{-\iu T_k \vartheta}\, \ldmax^N(T_k) \,,
\label{eq:faDensity}
\end{equation}
which is usually a good approximation to $\rho_{\osc}{(\vartheta)}$ already for a relatively small cut off parameter $\kc$.  
Particularly, in the near integrable regime,   a good approximation is obtained by just a few leading modes, see figure~\ref{fig:ApproximatedFourier1}. This is reflected  in a very pronounced oscillating structure of $\rho_{\osc}{(\vartheta)}$. In the chaotic regime there are no large gaps in the spectrum of the eigenvalues   \(\ldmax(T)\). This leads to a much more uniform density of eigenvalues, with a larger  cut off parameter $\kc$ necessary to resolve the fine structure of $\rho_{\osc}{(\vartheta)}$, compare figure~\ref{fig:ApproximatedFourier2}.

\subsection{Small Parameters Limit}
\label{smaPar}
\begin{figure}
\includegraphics[width=0.45\textwidth]{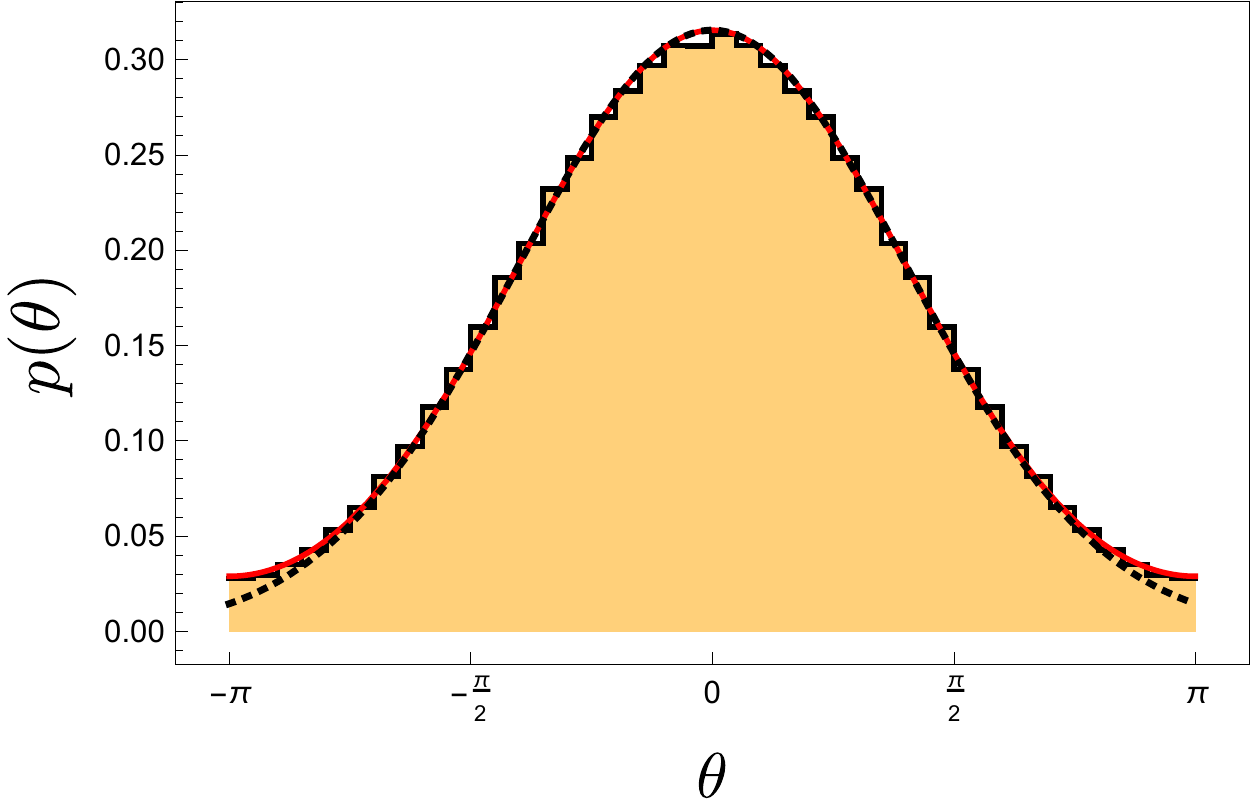}
\hfill
\includegraphics[width=0.45\textwidth]{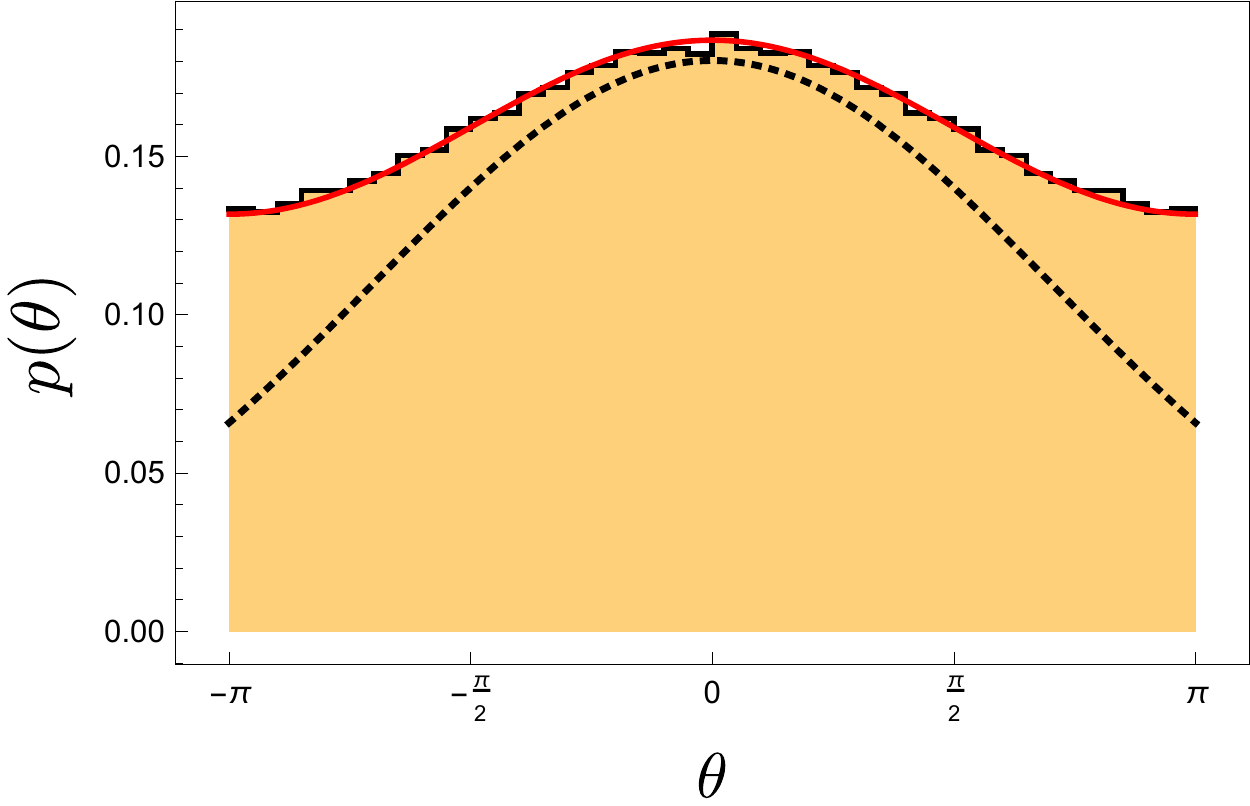}
\caption{Histogram of the quasi energies for \(J\teq b \teq 0.2\) (left) and \(J\teq b \teq 0.35\) (right), both for \(\varphi\teq \pi/2\) and \(N\teq 20\). The red (continuous) curve is given by the Gaussian approximation \eqref{eq:gausDensity}. The dotted black curve shows, for comparison, the contributions from \(n\teq 0\) to \eqref{eq:gausDensity}, only.}
\label{pic:gaussDense1}
\end{figure}

As an intermediate step between the discrete case and continues dynamics we look at the limit of small parameter \(J\) and \(\bm b\).
For  $J=0,\bm b = 0$  the (rescaled) dual operator 
\(\widetilde{U}\)
is a rank one projection operator. This suggests  that for small parameters  its eigenvalues can be evaluated 
by perturbation theory. To this end we introduce a small parameter \(\eta\) which rescales the parameters of our KIC model, \ie $J=\eta J_0,\bm b=\eta\bm b_0$.
To second order in $\eta$ the largest eigenvalue of $\widetilde{U}$  is given by
$ \tilde{\lambda}_1(\eta)= 2+ \eta \tilde{\lambda}_1'(0)+\tilde{\lambda}_1''(0) \eta^2/2 $, while  the rest of the eigenvalues $\tilde{\lambda}_i$ ($i> 1$) are of order  $O(\eta)$.  
Exploiting that these $\tilde{\lambda}_i(0)=0$ for $i> 1$ the  respective derivatives can be expressed through the traces,
\begin{equation}
 \partial_\eta \trace\widetilde{U}_T^2 \bigg|_{\eta=0}
 =2\tilde{\lambda}_1' \bigg|_{\eta=0}\,,
\end{equation}
\begin{equation}
 \partial^2_\eta\trace\widetilde{U}_T^3  \bigg|_{\eta=0} = \left(3\tilde{\lambda}_1'' \tilde{\lambda}_1^2 +6 \tilde{\lambda}_1'^{\;2} \tilde{\lambda}_1^2 \right) \bigg|_{\eta=0}
 \,. \label{feweigenvalues}
\end{equation}
At this point we can inverse the duality relation, leading to 
\begin{equation}
 \partial^k_\eta\trace\widetilde{U}_T^p\bigg|_{\eta=0}= \partial^k_\eta\trace\left(e^{-i \HI^{(p)} }e^{-i  \HK^{(p)} }\right)^T\bigg|_{\eta=0}\,,
\end{equation}
where $\HI^{(p)}, \HK^{(p)}$ are now $p$-spin Hamiltonians. Taking the derivatives and substituting $\eta=0$ this yields
\begin{equation}
\tilde{\lambda}_1'(0)=0
\quad\text{and} \qquad 12 \tilde{\lambda}_1''(0)=-T^2\trace (H^{(3)})^2,
\end{equation}
where $\hat{H}^{(3)}=\HI^{(3)}+\HK^{(3)}$ is  the sum of the kicked and Ising Hamiltonians for a $3$ particle spin chain. For the largest eigenvalue we therefore obtain by  (\ref{feweigenvalues}) 
\begin{equation}
\fl
\tilde{\lambda}_1(\eta)=2\left(1-\frac{T^2}{48}\trace (H^{(3)})^2\right)+O(\eta^3)= 2 \exp\left(-\frac{T^2\trace (H^{(3)})^2}{48}\right)+O(\eta^3)\,.\label{largesteig}
\end{equation}
The  trace in the exponent can easily be evaluated for an arbitrary number of spins $p$, see \cite{Atas}, leading to 
\begin{equation}
\frac{2^{-p}}{p}
\trace (H^{(p)})^2 =J^2+\bm b^2\,.
\end{equation}
Substituting  (\ref{largesteig}) into   (\ref{density_basic}) we can approximate  the density of eigenstates  by the sum
\begin{equation}
\rho{(\theta)}\approx \frac{1}{\pi}
\sum_{T=-\infty}^{+\infty}
\exp{\left(-\iu T\theta-\frac{1}{4} T^2\sigma^2\right)}\,,
\label{eq:preliminaryDensity}
\end{equation}
where \(\sigma^2= 2N (J^2+\bm b^2) \).
Due to the Poisson summation formula we finally obtain
\begin{equation}
\rho{(\theta)}\approx \frac{1}{\sigma \sqrt{\pi}}
\sum_{n=-\infty}^{+\infty}
\exp{\left(-\frac{(\theta+2\pi n)^2}{\sigma^2}\right)}\,,
\label{eq:gausDensity}
\end{equation}
which is  a periodized sum of Gaussians with widths given by \(\sigma\). Figure \ref{pic:gaussDense1} shows that this Gaussian approximation nicely agrees with the observed density.

\subsection{Continuous Limit}
\label{cont}

So far, we discussed the system in discrete time, where discontinuous kicks give the transition from one state in time to the next.
Below we show that the duality relation (\ref{duality}) can  also be used to evaluate  the eigenvalue density of   the rescaled Hamiltonian $\hat{H}=(\HK+\HI)/\sqrt{N}$ in the limit of large $N$.
The main idea here is to split the continues time  evolution into small steps  and approximate $\eu^{\iu \hat{H} t}$ by the  propagator of the KIC model.  

We follow the standard procedure and relate the smoothed density  function $\rho_\eps( E)$ to the traces of the evolution operator,  
\begin{equation}
\rho_\eps( E)
=
-\frac{1}{ 2^N \pi} \mbox{Im}\sum_{n=1}^{2^N}\frac{1}{E-E_n+\iu\eps}
=
\frac{1}{ 2^N \pi}\mbox{Im}\,\iu\int\limits_0^\infty\! \mathrm{d}t  \trace e^{-\iu t(H-E-i\eps)},
\end{equation}
where  $\eps$ is a small positive increment of the order $N^0$. 
Taking first the limit $N\to \infty$ and then  sending $\eps$ to zero yields the smoothed density of states $\rho( E)$.

Since most of the eigenvalues of $H$ are confined to the finite region around zero, it is sufficient to keep the traces $\trace e^{-\iu t H}$  under control   up to times  of the order $N^{\nu}$, where $\nu$ is an arbitrary positive number. To evaluate these traces  
we divide  the time  $t$ into a large number $M$ of small intervals  
\begin{equation}
\fl
 2^{-N} \trace\left(e^{-\iu tH}\right) = 2^{-N} \trace\left(e^{-\iu \tau \HI } e^{-\iu \tau  \HK }\right)^M+ 2^N O(t^2/ M), \quad t=\sqrt{N}M\tau\,.
\end{equation}
Due to the duality relation we have
\begin{equation}
\trace\left(e^{-\iu\tau \HI} e^{-\iu\tau  \HK}\right)^M=\trace \left(\widetilde{U}_M(\tau) \right)^N = \sum_{k=1}^{2^M} \tilde{\lambda}_k^N(\tau)\,,\label{densityduality}
\end{equation}
where $\tilde{\lambda}_k$ are the eigenvalues of the dual  $2^M\times 2^M$ matrix $\widetilde{U}_M(\tau)$. Its  parameters are determined by (\ref{dualmatrix}).
In the special case of $\tau=0$,  $\widetilde{U}_M(\tau)$ has only one non-zero eigenvalue $\tilde{\lambda}_1=2$. At small $\tau$ resp. large $M$ we can use the perturbative expansion
\begin{equation}
\tilde{\lambda}_1(\tau)=2+\sum_{k=1}^\infty c_k\tau^k, \qquad \tilde{\lambda}_{n>1}(\tau)=O(\tau)\,,
\end{equation}
to evaluate (\ref{densityduality}). For the smallest $k$'s, 
the constants $c_k$  can easily be  found by relating them to the derivatives of $\widetilde{U}_M(\tau)$, as in the previous subsection. To leading order in $1/N$ this gives  
\begin{equation}
\tilde{\lambda}_1(\tau)= 2 \exp\left(-\frac{t^2\langle H^2\rangle}{2N}\right)+O(1/N^2)\,,
\end{equation}
 where the average $\langle H^2\rangle=2^{-N}\trace H^2$ is independent of $N$. After insertion of $\lambda_1(\tau)$ into (\ref{densityduality}),  taking the limits $N,M\to\infty$ and sending $\eps$ to zero we obtain 
\begin{equation}
\fl
\rho( E) =
\mbox{Im}\frac{\iu}{\pi  }\int\limits_0^\infty\! \mathrm{d}t  \exp\left(-\frac{t^2\langle H^2\rangle}{2}+\iu tE \right) =
 \frac{1}{\sqrt{2\pi\langle H^2\rangle}}\exp\left(-\frac{E^2}{2\langle H^2\rangle}\right)\,.
\end{equation}
This is precisely the result which was  obtained in  \cite{Atas,keat1,keat2}  by different methods.

\section{Spectral Form Factor}\label{section4}

In section \ref{sec:FoFaDuaRel} we recapitulate the predictions for the form factor \(K_2(T)\) based on Random Matrix Theory (RMT), for an overview see \cite{Guhr}, valid for large times \(T\) in the chaotic regime.
In section \ref{sec:shortT} we use the duality relation to obtain analytic expressions for the short time form factor at arbitrary particle numbers.
For longer times we give approximations for the spectral form factor, considering separately the integrable and the chaotic case in sections \ref{sec:longT} and \ref{sec:longTchaos}.

\subsection{Form Factor via RMT}
\label{sec:FoFaDuaRel}
The spectral form factor is defined as the  Fourier transform of the two-point spectral correlation function, see \eg \cite{haake3,prosen2007}. Using the duality relation it can be represented in the form
\begin{equation}
 K_2(T)=\frac{1}{2^N}|\trace{{U}_N^T}|^2=\frac{1}{2^N}|\trace {\tilde{U}_T}^N|^2\,.
 \end{equation}
  By its very definition    $ K_2(T)$ carries information about the correlations between eigenphases $\vartheta_n$ on the scales of  $1/T$. If the system is in the chaotic regime, for large times $T$ comparable with the Heisenberg time of the system $2^N$,
one expects that $ K_2(T)$ has the  
universal form provided by the relevant RMT ensemble. 
  For the KIC the total spectrum  can be split  into $(N-2)/2$ (resp. $(N-1)/2$) doubly degenerate   and $4$ (resp. $2$)  non-degenerate subspectra if  $N$ is even  (odd), see \cite{prosen2007}. Each sector has orthogonal symmetry, and as \(U_N\) is unitary, this implies spectral statistics comparable to the circular orthogonal matrix ensembles (COE). Indeed, in the KIC such universal spectral correlations can be observed  in the chaotic parameter regime, see \cite{prosen2007} for details.   Furthermore, in the \mbox{large-$T$} limit all subspectra can be considered as independent. Therefore the expected total form factor in the chaotic regime is
\begin{equation}
K_2(T)= 2K^{\text{(COE)}}_2(NT/2^N)\,,
\label{eq:rmtFormFac}
\end{equation}
with the universal COE form factor
\begin{eqnarray}
K^{\text{(COE)}}_2(\tau)=
\left\{
\begin{array}{lr}
2|\tau|-|\tau|\log{\left(1 +2|\tau| \right)}
& \text{for } |\tau| \leq 1
\\
2-|\tau|\log{\frac{2|\tau|+1}{2|\tau|-1}}
& \text{for } |\tau| > 1
\end{array}
\right.\,.
\label{RMTFormFactor}
\end{eqnarray}
We assumed that $N$ is large, such that the four (resp. two) special  sectors can be ignored.

\subsection{Short Times}
\label{sec:shortT}

In \cite{prosen2007} Pineda and Prosen studied the spectral form factor numerically also at short times for up to 20 particles. They found a behavior not consistent with RMT. 

\begin{figure}[bhtp]
\centering
\includegraphics[width=0.45\textwidth]{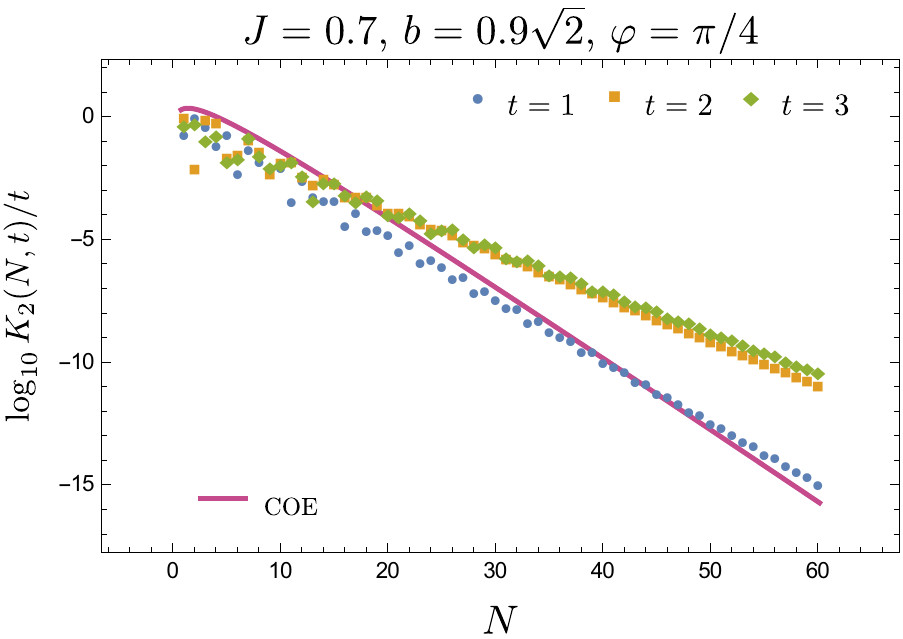}
\hfill
\includegraphics[width=0.45\textwidth]{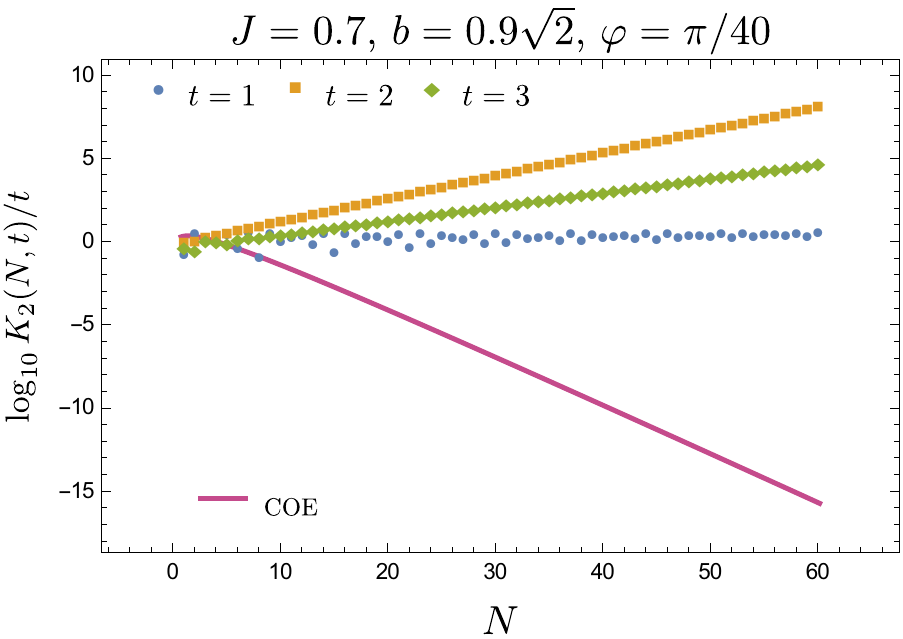}
\\
\includegraphics[width=0.45\textwidth]{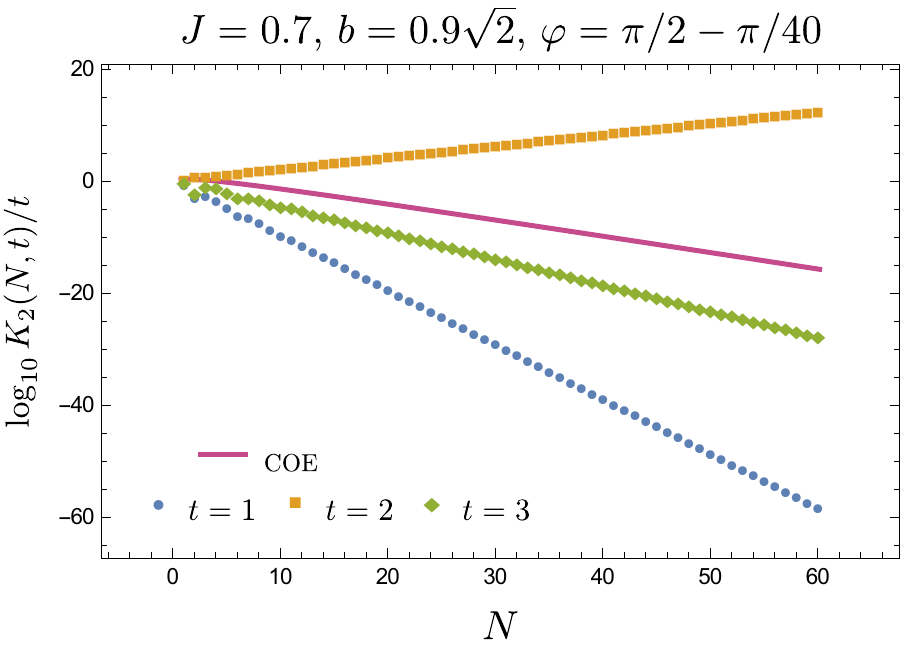}
\caption{Logarithm of the form factor \(K_2(N,t)\) as a function of \(N\) for the total spectrum and short times \(t\). As comparison we plot the COE result \(\log_{10}{(N 2^{2-N})}\) (continuous line). The upper left panel shows the chaotic regime, where the form factor follows RMT predictions more closely, for details see text. The other two panels consider parameters close to the integrable points and depict strong deviations.}
\label{pic:ffdec1alt}
\end{figure}

\begin{figure}[bhtp]
\includegraphics[width=0.3\textwidth]{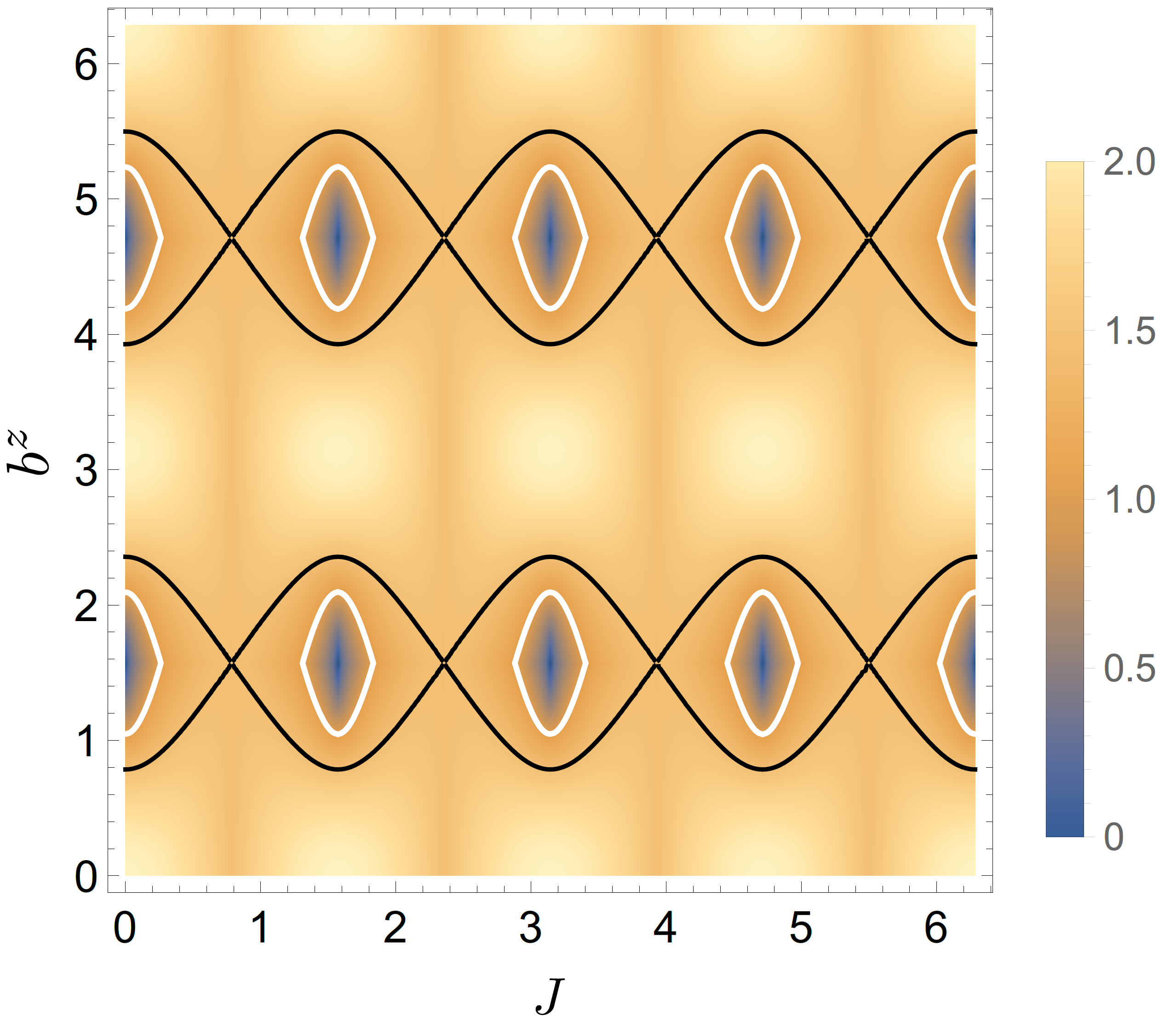}
\hfill
\includegraphics[width=0.3\textwidth]{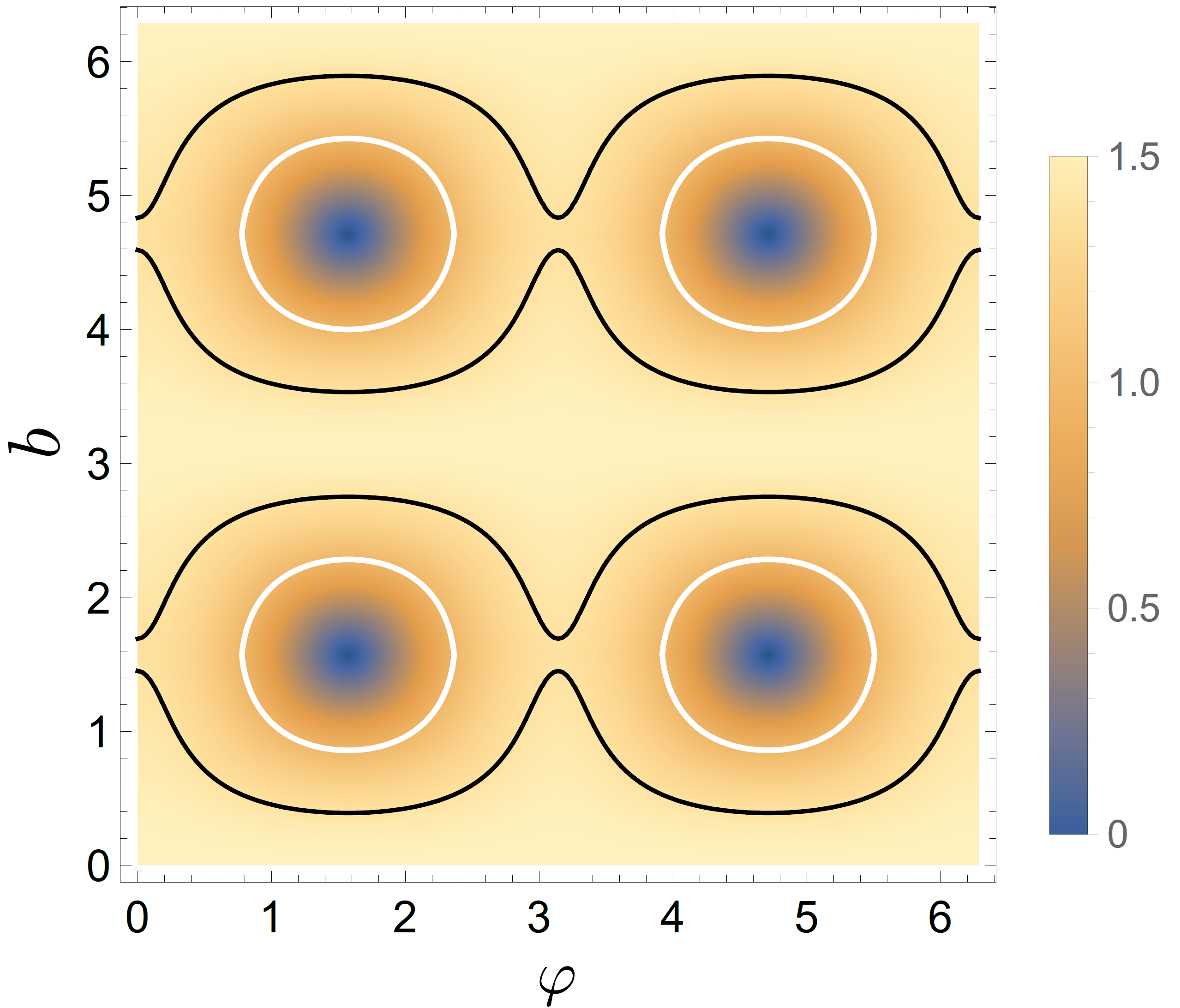}
\hfill
\includegraphics[width=0.3\textwidth]{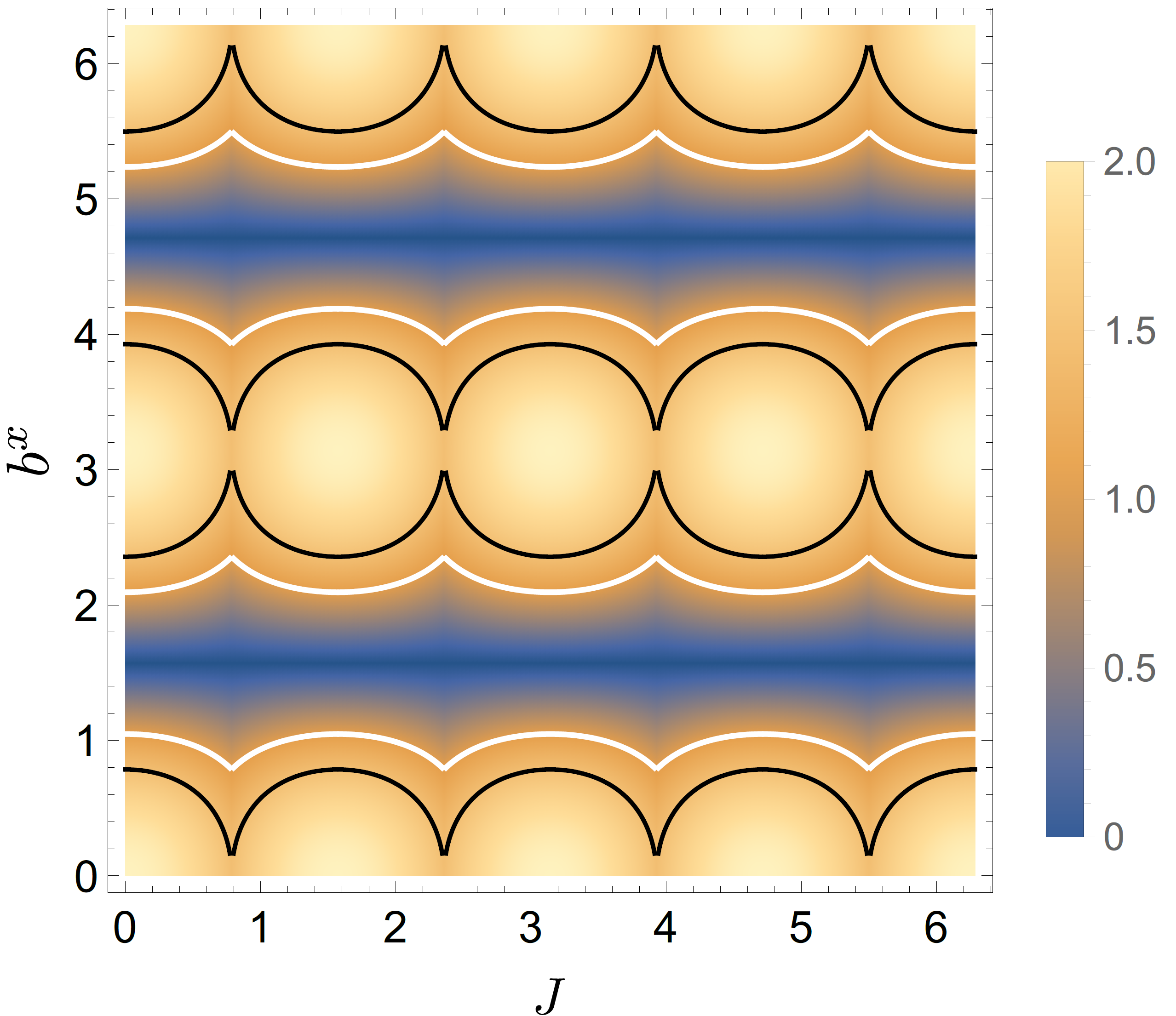}
\caption{Absolute value of \(\ldmax\) for different parameters of \(J\) and \(b\) in the trivially integrable case (\(\varphi\teq 0\)) (left) and the non-trivially integrable case (\(\varphi\teq \pi/2\)) (right). The contour lines indicate the special cases of \(\ldmax\teq 1\) (white) or \(\sqrt{2}\) (black) respectively. The middle panel shows the transition from one integrable domain to another and back when varying \(\varphi\). The coupling is fixed to \(J \teq 0.7\).}
\label{pic:lambdaBehaviourRegions}
\end{figure}

In this context we can apply the duality relation \eqref{duality} to analytically compute 
the spectral form factor for arbitrary numbers of particles in the short time regime.
For $T\leq 3$, the spectrum of the dual   operator \(\tilde{U}_T\)  can be calculated  analytically and  a closed expression for $ K_2(T)$ can be provided. In particular, for the one time step form factor we have
\begin{equation}
K_2(1)=
\frac{1}{2^N}|\tilde{\lambda}_+^N + \tilde{\lambda}_-^N|^2\,,\label{StepOneFormFactor}
\end{equation}
where $\tilde{\lambda}_\pm$ are the two eigenvalues of
\begin{equation}
\tilde{U}_1 = g
\left(
\begin{array}{cc}
 \eu^{-\iu K} (\cos{\tilde{b}}+\iu \cos{\tilde{\varphi}} \sin{\tilde{b}}) & -\iu \eu^{-\iu K} \sin{\tilde{b}} \sin{\tilde{\varphi}} \\
 -\iu \eu^{-\iu K} \sin{\tilde{b}} \sin{\tilde{\varphi}} & \eu^{-\iu K} (\cos{\tilde{b}}-\iu \cos{\tilde{\varphi}} \sin{\tilde{b}}) \\
\end{array}
\right)
\,,
\end{equation}
given by
\begin{equation}\label{eigenv}
\tilde{\lambda}_\pm=\cos{b}\,\eu^{-\iu J}
\pm \sqrt{\eu^{2\iu J}-\sin^2{b}\,
\left( \eu^{2\iu J} \sin^2\varphi
+\eu^{-2\iu J} \cos^2\varphi\right)}\,.
\end{equation}
For small $N$,  the function  $K_2(1)$ strongly  fluctuates in dependence of $N$, see figure~\ref{pic:ffdec1alt}.  
However, in the limiting case of \(N\gg 1\)  only the eigenvalue with the larger absolute magnitude, denoted by \(\ldmax\), will contribute to the form factor. Asymptotically, we therefore find
\begin{equation}
\frac{1}{N}\log K_2(1)\sim \log \left(|\ldmax|^2/2\right).
\end{equation}
To compare this result to the RMT universal form factor, where, for $N\rightarrow\infty$,  \(\log K_2(1)\sim -N \log 2 \) is expected, we  distinguish 3 asymptotically different behaviors: 1) \(|\ldmax| <1 \) where the form factor decays faster than the corresponding COE, 2) \( 1<|\ldmax|<\sqrt{2}\) where the decay is slower and 3) \(|\ldmax|>\sqrt{2}\), where \(K_2(1)\) grows with \(N\).
Figure \ref{pic:lambdaBehaviourRegions} shows the boundary lines for both integrable regimes \(\varphi\teq 0\) and \(\varphi\teq \pi/2\) as well as the transition from one to the other.

Similarly,  for the two step form factor we have
\begin{equation}
K_2(2)=
\frac{1}{2^N}|\Tr{U^2_N}|^2=
\frac{1}{2^N}|\Tr{\tilde{U}_2^N}|^2=
\frac{1}{2^N}|\tilde{\lambda}_0^N + \tilde{\lambda}_1^N+ \tilde{\lambda}_2^N+ \tilde{\lambda}_3^N|^2\,,
\end{equation}
where $\tilde{\lambda}_k$ are the four eigenvalues of  \(\tilde{U}_2\), given by
\begin{equation}
\tilde{U}_2=g^2 \eu^{2\eta}
\left(
\begin{array}{cccc}
 \eu^{2 \iu (h-J-K)} & \eu^{\iu h-2 \iu K} & \eu^{\iu h-2 \iu K} & \eu^{2 \iu J-2 \iu K } \\
 \eu^{\iu h+2 \iu K} & \eu^{2 (-\iu J+\iu K )} & \eu^{2 \iu (J+K)} & \eu^{-\iu h+2 \iu K} \\
 \eu^{\iu h+2 \iu K} & \eu^{2 \iu (J+K)} & \eu^{2 (-\iu J+\iu K)} & \eu^{-\iu h+2 \iu K} \\
 \eu^{2 \iu J-2 \iu K} & \eu^{-\iu (h+2 K)} & \eu^{-\iu (h+2 K)} & \eu^{-2 \iu (h+J+K)} \\
\end{array}
\right)
\,.
\end{equation}
Using the translational symmetry of the system it is easy to
see that  the fully antisymmetric state \(\left| +1,\, -1  \right\rangle - \left| -1,\,+1 \right\rangle \) is also an eigenstate of \(\tilde{U}_2\) with the eigenvalue $\tilde{\lambda}_0 \teq g^2e^{2iK} \teq -2\iu \sin^2{\varphi}\,\sin^2{b}\sin{2J}$. The rest of the dual spectrum can be found as eigenvalues of the remaining $3\times3$ matrix,
\begin{equation}
\tilde{U_2}^\text{(sym)}=g^2 \eu^{2\eta}
\left(\begin{array}{ccc}
\eu^{-2\iu (J+K-h)} & \eu^{+2\iu (J-K+h)} & \sqrt{2}\eu^{-2\iu (K-h)} \\
\eu^{+2\iu (J-K-h)} & \eu^{-2\iu (J+K+h)} & \sqrt{2}\eu^{-2\iu (K+h)} \\
\sqrt{2}\eu^{+2\iu K} & \sqrt{2}\eu^{+2\iu K} & 2\eu^{2\iu K}\cos{2J} \\
\end{array}\right)\,.
\end{equation}
In the case of \(K_2(3)\) the dual matrix can be decomposed into one \(4\!\times\!4\) block and two \(2\!\times\!2\) blocks. Figure\  \ref{pic:ffdec1alt} shows the resulting form factors for all three time steps together with the RMT prediction \eqref{eq:rmtFormFac}. As stated in the beginning of this section, \cite{prosen2007} pointed out that the (short-time) form factor in the chaotic case stays systematically and significantly below the expected RMT results. As seen in the figure this behavior changes with increasing \(N\), as only \(|\ldmax|\) determines the asymptotic result.

\subsection{Long  Times --- Integrable Regimes}
\label{sec:longT}
In the integrable regimes the behavior of the spectral form factor is expected to differ from the RMT result, even for long times. For the trivially integrable case $\varphi \teq 0$ the form factor can be easily calculated for an arbitrary time. Because of 
\begin{equation}
 K_2(T)=\frac{1}{2^N}\left|\trace e^{-\iu T\HI }e^{-\iu T\HK }\right|^2\,,
\end{equation}
it is given by the same expression (\ref{eigenv}) as  the one-step form factor for $\varphi\teq 0$,   
\begin{eqnarray}
K_2(T)&=\frac{1}{2^N}|\tilde{\lambda}_+^N + \tilde{\lambda}_-^N|^2\,,
\\
\tilde{\lambda}_\pm(T)&={\eu}^{-\iu JT}\left(\cos{(bT)}
\pm \sqrt{\eu^{\iu 4JT}-\sin^2{(bT)}}\right)\,,
\end{eqnarray}
 with rescaled parameters $b\to bT$, $J\to JT$.
 
\begin{figure}[bhtp]
\centering
\includegraphics[width=0.45\textwidth]{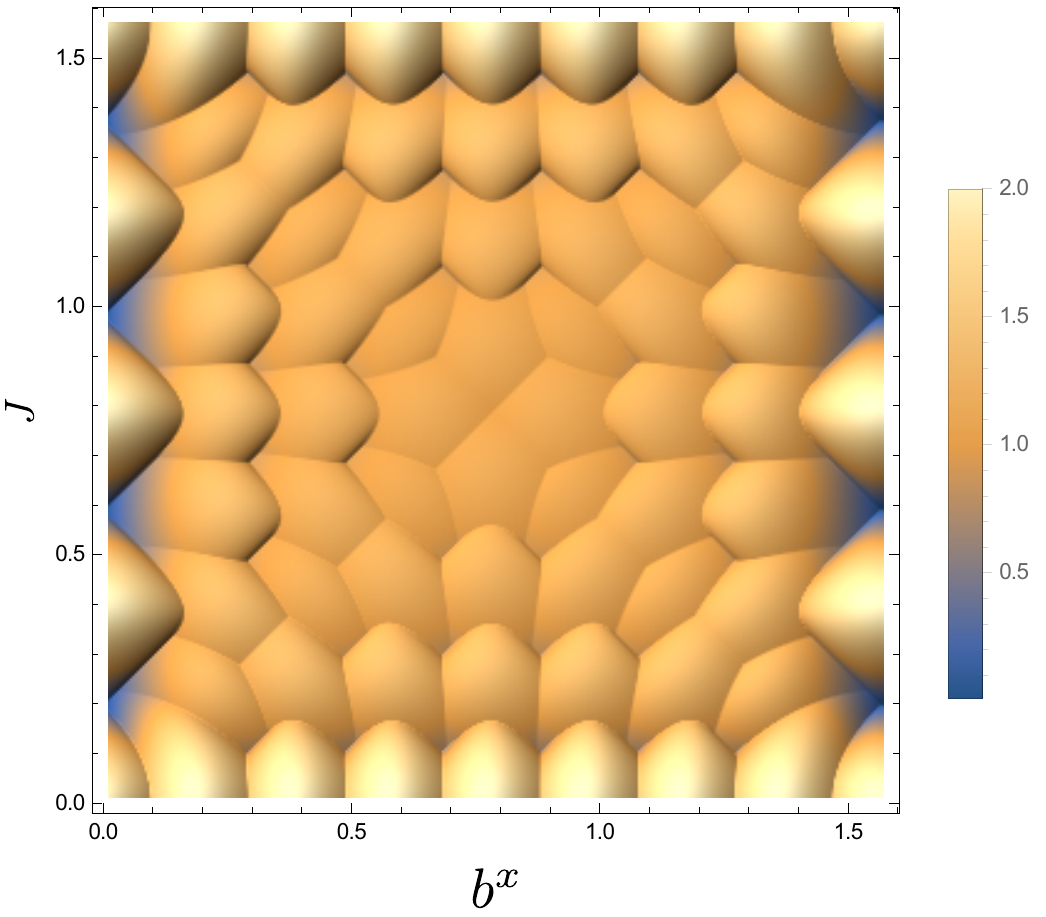}
\hfill
\includegraphics[width=0.45\textwidth]{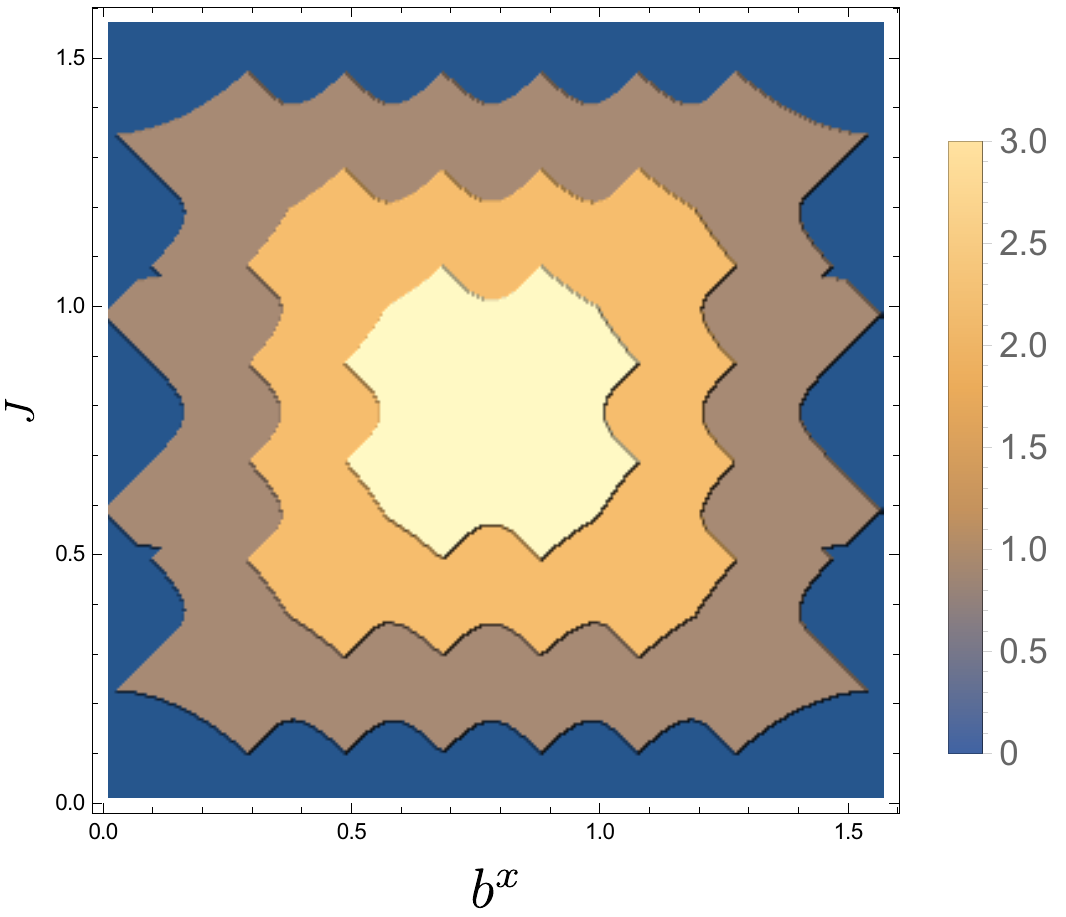}
\caption{Left: Absolute value of the largest dual eigenvalue \(|\ldmax|\) over the \((J,\,b^x)\) plane in the non-trivially integrable regime for \(T\teq 8\). Right: Same dependence for \(n\), also for  \(T\teq 8\). The number of layers grows with \(T\) but retains the pyramidal form, compare \cite{prevPap}.}
\label{fig:NTldmax}
\end{figure}

Although in the non-trivially integrable regime $ K_2(T)$ can be obtained explicitly, the resulting expression is quite cumbersome and its asymptotic analysis at $N\to\infty$ is difficult. It is  instructive  to use for this purpose the duality approach.  
In the dual picture the form factor for large particle numbers  is strongly dominated by the eigenvalues with the largest  absolute magnitude \(\ldmax\). Restricting thus the sum to the outer circle of \(\tilde{U}_T\), yields
\begin{equation}
K_2(T)=\frac{1}{2^N}\left<\left|\sum_{i=0}^{2^T}\,|\tilde{\lambda}_i|^N\eu^{\iu N\tilde{\vartheta}_i} \right|^2\right>
\approx \frac{|\ldmax|^{2N}}{2^N} \left<\left| \sum_{\substack{ \text{outer} \\ \text{circle}}}\,\eu^{\iu N\tilde{\vartheta}_i} \right|^2\right>
\,,
\label{eq:outerApprox}
\end{equation}
where the integer parameter \(n\teq n(T)\) controls the number of eigenvalues and their degeneracies on said circle. A more in depth discussion of the spectrum of this operator can be found in \cite{prevPap}.
 To smoothen the large fluctuations of the form-factor we take an average $\left<\cdots \right>$ over the parameters $J, b^x$ of the spin chain, such that both   $n$ and $\ldmax$ are kept fixed. The dependence of both on the parameters is shown in figure \ref{fig:NTldmax}.

\begin{figure}
\centering
\includegraphics[width=0.42\textwidth]{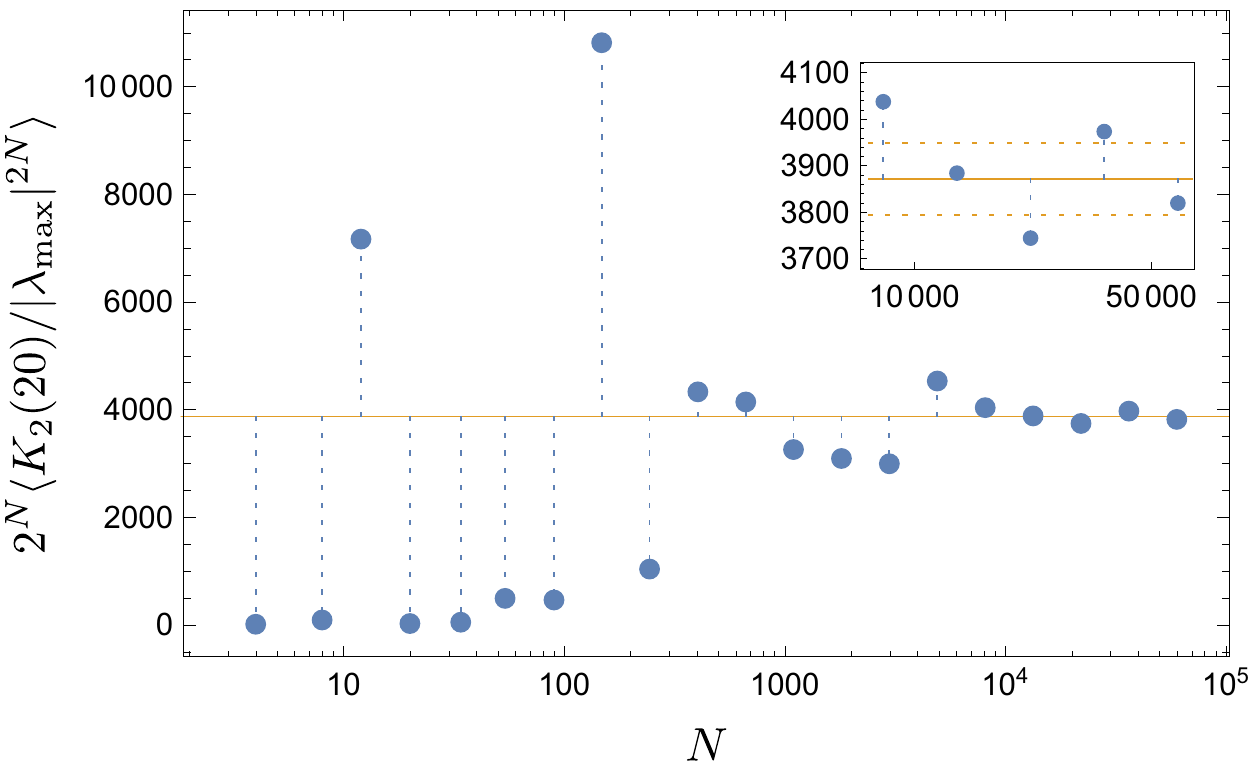}
\hfill
\includegraphics[width=0.45\textwidth]{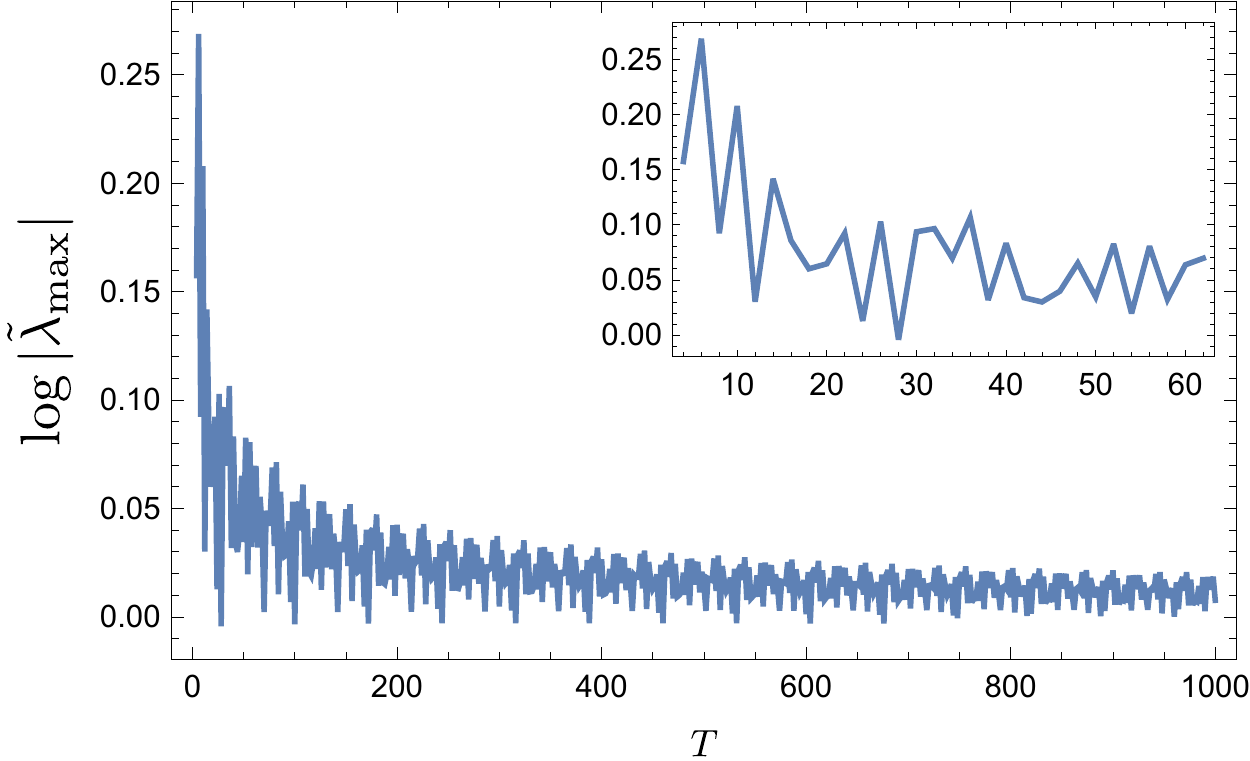}
\caption{To the left: convergences of the approximation \eqref{eq:diagapprox}. The blue dots correspond to the average form factor for \(T\teq 20\) and \(n\teq 5\).  The average is taken over \(136\, 000\) points along a closed \(|\ldmax|\) contour. The straight line shows the expected result of \eqref{eq:diagapprox} for the given \(n\). The inset shows the remaining deviations for large \(N\), dashed lines indicate \(2\%\) deviation.
To the right: \(|\ldmax|\) in dependence of \(T\) for \(J\teq 0.48,\,b^x\teq 1.2,\,N\teq 10^4\). The inset magnifies the initial fluctuations for short times.}
\label{fig:NTdualLargeTSum}
\end{figure}

As follows from \eqref{eq:outerApprox} the behavior of $K_2(T)$ as a function of $T$  is controlled in this large $N$ regime  by $\ldmax(T)$ and $n(T)$.
Assuming that $N$ is larger than the number of distinctive eigenvalues at the outer circle  we can apply the diagonal approximation to  the right-hand side of \eqref{eq:outerApprox}.    The resulting  sum  depends only   on the degeneracies \(d_i\) and multiplicities \(m_i\) of the eigenvalues $\ldmax\eu^{\iu \tilde{\vartheta}_i} $ and can be evaluated explicitly, resulting in 
\begin{equation}
\label{eq:diagapprox}
\left<\left| \sum_{i=1}^{2^{2n-1}}\,\eu^{\iu N\tilde{\vartheta}_i} \right|^2\right>
\approx
\sum_{i=0}^{\lfloor n/2 \rfloor}\,m_i\,d_i^2
=
2^{n-1}\Big(3^n+(-1)^n \Big)
\,,
\end{equation}
where the respective summands are given by
\begin{equation}
d_i=2^{2i}
\qquad\text{and}\qquad
m_i=2^{n-2i}\binom{n}{2i}
\,.
\label{eq:degDef}
\end{equation}
Up to  jumps of the order of unity \(n\) can be approximated  by a ``smooth'' linear  function of $T$,
\begin{equation}
n(T)=\nu T +\mathcal{O}(1), \quad \nu(J,b^x)=\frac{2}{\pi}\min\{J,b^x,\frac{\pi}{2}-J,\frac{\pi}{2}-b^x\}\,,\label{eq:outerApprox2}
\end{equation}
valid for the default parameter range \(J,b^x\in [0,\pi/2]\).
To verify  the validity of this diagonal approximation we  perform a numerical average  of \eqref{eq:diagapprox} in dependence of $N$, the result is shown in figure \ref{fig:NTdualLargeTSum}.
To complete the analysis we still need to evaluate $\ldmax(T)$. Due to the results from \cite{prevPap} it can be written as
\begin{equation}
\label{eq:ldmax}
\log{|\ldmax|}=
\sum_{k=1}^{\lfloor T/2 \rfloor} \Theta(\beta(k))
\log{\left(|\alpha(k)|+\sqrt{\beta(k)} \right)}
+T \log{g}\,,
\end{equation}
where \(\Theta\) is a Heaviside-Step-Function and the parameters are given by
\begin{eqnarray}
\label{eq:alphabeta}
\alpha(k)=-\frac{1}{g^2}\left(\cos{2b^x}\cos{2J}-\cos{2\vartheta(k)}\right)
\,, \\
\nonumber
\beta(k)=\alpha(k)^2-1\,,
\end{eqnarray}
with
\begin{equation}
g^2=\sin{2b^x}\sin{2J}
\qquad\text{and}\qquad
\vartheta(k)=\frac{2\pi}{T}\left(2k+1\right)\,.
\end{equation}
From the numerically obtained graph, presented in figure \ref{fig:NTdualLargeTSum}, it can be deduced that \(\log|\ldmax(T)|\) saturates to a fixed value \(C_0\) dressed by time dependent fluctuations \(\xi(T)\).
Incorporating this into \eqref{eq:diagapprox} we obtain for the form factor
\begin{equation}
 K_2 (T)\sim \exp\Big( (C_0 +  \xi(T))\; N+ T \nu \log 6\Big)\,. 
\end{equation}
This expression suggests  that for short times the form factor grows exponentially with $T$. However, such an exponential growth would be  difficult 
to detect as it is masked by the large fluctuating term $\xi(T) N$.

\subsection{Long  Times  --- Chaotic Regime}
\label{sec:longTchaos}
It was observed in \cite{prosen2007}
that for the  KIC with  parameters $J, \bm b$  in the chaotic regime the   spectral statistics of $U_N$ obey the universal RMT predictions. For the spectral form factor this implies a linear growth with time,
\begin{equation}
 \K(T,N)=2^N K_2(T)=4 NT\left( 1+O\left(\frac{NT}{2^N}\right)\right)
 \,. \label{RMTformfac}
\end{equation}
Strictly speaking, (\ref{RMTformfac}) should be expected    when $T$ is of the same order as the Heisenberg time $2^N/N$ for an individual sector. However, for a Hamiltonian systems with underlying chaotic classical dynamics the linear growth of $\K(T,N)$ persists up to much  shorter classical  time scales. 
From the semiclassical point of view this can be attributed to diagonal correlations between periodic orbits in the double sum representing \(K_2(T)\), see \cite{Berry}. While non diagonal correlations can arise only for time scales larger than the Ehrenfest time \cite{waltner,waltner2, pbrauer}.
Interestingly, 
similar considerations
give rise to the linear growth  for the KIC as well. To illustrate this we cast (\ref{partfunction}) into  a ``semiclassical'' form,
\begin{equation}
 Z(N,T)=\sum_{\bm\sigma}A_{\bm\sigma}\exp\left(-\iu S(\bm\sigma)\right), \label{semiclassical}
\end{equation}
where
\begin{eqnarray}
A_{\bm\sigma}=\exp\left[-\iu \sum_{n=1}^N \sum_{t=1}^T \left(K\sigma_{n,t} \sigma_{n,t+1} +\iu\eta\right) \right]\,,
\\
S(\bm\sigma)=\sum_{n=1}^N\sum_{t=1}^T J\sigma_{n,t}\sigma_{n+1,t}+h\sigma_{n,t}
\,.
\end{eqnarray}
Equation~(\ref{semiclassical}) is reminiscent of the Gutzwiller trace formula with  $S(\bm\sigma),\, A_{\bm\sigma}$  as the actions and stabilities of $2^{NT}$ ``periodic orbits'' which are labeled by all possible spin configurations  $\bm\sigma\teq\{\sigma_{n,t}\teq\pm 1\}$. Applying the standard diagonal approximation, see \cite{Berry}, yields 
\begin{equation}
 \K(T,N)/4NT=\sum_{\bm\sigma}\left|A_{\bm\sigma}\right|^2.
 \end{equation}
Substituting therein the amplitudes $A_{\bm\sigma}$ we obtain
\begin{eqnarray}
\K(T,N)/4NT &=& \left(\sum_{\bm\sigma} \exp \left( -2 K_0\sum_{t=1}^T \sigma_{t} \sigma_{t+1}-2\eta_0 T \right) \right)^N \\
\nonumber
&=& \left(1+(\tanh 2K_0)^T \right)^N
\,,
\end{eqnarray}
where we introduced $\eta=\iu\pi/4-\eta_0$ and $K=\pi/4-\iu K_0$, \cf \eqref{eq:khhparam}. The sum over all configurations \(\bm\sigma\) now covers only individual particles for fixed \(n\), the summation over different particles is replaced by the \(N\)-th power.
In the limit of large $N$ and $T$ the right hand side  tends to unity leading to the  expected RMT result provided that \(N<\coth^T{2K_0}\).

Alternatively, such a linear growth can be seen as a symmetry factor from the dual perspective. Indeed, we have 
\begin{equation}
 \K(T,N)=\trace(\tilde{U}_T)^N \sim T
\end{equation}
since $\tilde U_T$ has $T$ different subspectra due to translational invariance, which  for large $N$  can be thought of as uncorrelated.


\section{Conclusion and Outlook}\label{section5}
The Hilbert space dimension of a many-body system scales exponentially with the number of particles, this implies that brute-force methods are only applicable for very small system sizes.
For a kicked Ising spin chain we studied traces of the propagator for a large number of particles \(N\).
Due to the large dimension of the system's Floquet operator \(U_N\) this regime has previously not been easily accessible.
Our approach is based on a trace duality between \(U_N\) and a dual matrix \(\tilde{U}_T\), reminiscent of an ``evolution operator'' in particle direction, whose dimension \(2^T\!\times\! 2^T\) is given by the number of time steps \(T\) used for the actual evolution. 
This matrix has the important benefit that it allows
statements about the eigenvalue density of the Floquet operator and the spectral form factor
in the limit of many particles and small times, as the dimension of this dual matrix is small in this case. 
In our model it turned out that \(\tilde{U}_T\) is non-unitary and features a rich structure, which depends on the parameter regime considered.

Furthermore, we exploited the non-unitarity for large \(N\) asymptotics as in those cases only the largest eigenvalues of  \(\tilde{U}_T\) play a significant role.
Following this line of reasoning we gave approximations to the smoothed density of states where we found abrupt transitions in the dominant oscillation frequency.
We also investigated the limit of small system parameters \((J,\,\bm{b})\) which allowed us to study properties of time continuous spin chains. Here, the trace duality provides an easy to use mechanism to derive the 
density profile for the eigenvalues of the Hamiltonian.

For the form factor we can, due to the reduced dimensionality of \(\tilde{U}_T\), give explicit expressions for short times \(1\leq T \leq 3\) but arbitrary \(N\). This shines some light on previously raised questions regarding non RMT behavior of the form factor in this limit, see \cite{prosen2007}.
In the integrable regime exact expressions are, in principle, known as the model can be mapped on a solvable classical two dimensional Ising model or free fermions. 
Still, we can use the duality relation to easily obtain asymptotic and approximate results based on the largest eigenvalues. In this case the result not only depends on the eigenvalues magnitude, but also on the degeneracy which changes abruptly when varying the system parameters.
For the chaotic case the known linear growth in time for the form factor was derived building on periodic orbit correlations, but it can also be understood from the structural decomposition of \(\tilde{U}_T\) into \(T\) sectors.

It would be of great interest to understand  the asymptotic behavior of  the   spectral form factor in the chaotic regime also for a large but finite $T$ when $N\to\infty$.
In this regime the eigenvalues of the unitary  KIC propagator obey the universal COE distribution on the scales of mean level spacing.  A natural and intriguing question  arises with respect to  the eigenvalues of the dual propagator: whether they  exhibit  universal behavior  which could be described by a non-unitary random matrix ensemble? This in turn  would imply  some kind of  ``universality'' in  the spectral correlations   at scales much larger  than conventional mean level spacing.

As a closing remark, let us point out that the duality relation remains valid also  in the case when  kick  and/or coupling parameters are chosen randomly.
In such a case this relation might proof a useful tool to study \eg many body localization.
Another possible area of application
 are Hamiltonian systems with continues time dynamics. Splitting the propagation into sufficiently small time steps the dynamics can be approximated 
 by kicks in discrete time.
This   opens a road  to apply  the duality relation to short time propagators of Hamiltonian systems as well.

 \end{document}